\documentclass[12pt]{article} 
\usepackage{amsmath}
\usepackage{amssymb} 
\usepackage{graphics} 

 \def\C{{ \mathbb C}} \def\R{{ \mathbb R}}     

    \def\tr{\operatorname{tr}}

 \def\tr{{\rm tr\, }} 
 \def\Tr{{\rm Tr\, }}

\def\id{\protect{{1 \kern-.28em {\rm l}}}}

\newcommand{\be}{\begin{equation}} \newcommand{\ee}{\end{equation}}
\newcommand{\bea}{\begin{eqnarray}} \newcommand{\eea}{\end{eqnarray}}
\newcommand{\beann}{\begin{eqnarray*}}
  \newcommand{\eeann}{\end{eqnarray*}}
\newcommand{\bfig}{\begin{figure}} \newcommand{\efig}{\end{figure}}

\newcommand{\ba}{\begin{array}}\newcommand{\ea}{\end{array}}

\newtheorem{Proposition}{Proposition}[section]

\newtheorem{Theorem}{Theorem}[section]
\newtheorem{Lemma}{Lemma}[section]
\newtheorem{Corrolary}{Corrolary}[section]

\newcommand{\bp}{\begin{Proposition}}
  \newcommand{\ep}{\end{Proposition}}
\newcommand{\bt}{\begin{Theorem}} \newcommand{\et}{\end{Theorem}}
\newcommand{\bl}{\begin{Lemma}} \newcommand{\el}{\end{Lemma}}
\newcommand{\bc}{\begin{Corrolary}} \newcommand{\ec}{\end{Corrolary}}

\def\None{{\cal N}\!\!=\!\!1}
\def\Ntwo{{\cal N}\!\!=\!\!2}

   \def\ep{\eps}
 \def\hn{{ \hat N}} \def\hq{{\hat Q}}  \def\hs{{ \hat S}} \def\ha{{\hat A}} 
\def\hx{\hat X}\def\hy{\hat Y}\def\hf{\hat f}
\def\cW{{\cal W}}

\def\erf#1{(\ref{#1})}

\begin{document}

\begin{titlepage}
\begin{flushright}
IFT-UAM/CSIC-04-31
\end{flushright}
\vspace{1cm}
\begin{center}{\Large\bf
Konishi Anomalies and Curves without Adjoints}\\ \vspace{1.5cm}
{\large 
K. Landsteiner}\\
\vspace{1cm}
{\it Insitituto de F\'\i sica Te\'orica,} \\
{\it C-XVI Universidad Auton\'oma de Madrid, }\\
{\it 28049 Madrid, Spain}\\
{\tt Karl.Landsteiner@uam.es} \\ \vspace{1cm}
\vspace{2cm}
\bf Abstract
\end{center}
\noindent 
Generalized Konishi anomaly relations in the chiral ring of $\None$
supersymmetric gauge theories with unitary gauge group and chiral
matter field in two-index tensor representations are derived. 
Contrary to previous investigations of related models we do not
include matter multiplets in the adjoint representation.
The corresponding curves turn out to be hyperelliptic.
We also point out equivalences to models with orthogonal
or symplectic gauge groups.
\end{titlepage}

\section{Introduction} 

Supersymmetric field theories are among the most interesting theoretical
laboratories to investigate gauge theories in the strongly coupled
regime. It is often possible to obtain
exact results on the non perturbative dynamics.
A recent prominent example is the usage of a generalized form of the
well-known Konishi anomaly \cite{Konishi:1984eu} to compute the exact
effective superpotential \cite{Cachazo:2002ry}, see also
\cite{Gorsky:2002uk}. This usage of generalized Konishi anomalies
was motivated by the the Dijkgraaf-Vafa conjecture \cite{DV}. It says
that the exact
effective superpotential of a confining
$\None$ supersymmetric gauge theory can be computed using a matrix model
whose action is given by the tree level superpotential of the field
theory. 
 
The prototypical example in this line of research is $\Ntwo$ supersymmetric
gauge theory softly broken to $\None$ by a superpotential for the chiral
multiplet in the adjoint representation. 
Let us briefly recall the construction of \cite{Cachazo:2002ry}. 
With the help of the generalized Konishi anomalies one finds algebraic
relations for the gauge invariant operators
\begin{equation}
  \label{eq:tandr}
  T(z) = \left\langle\tr \left( \frac{1}{z-\phi} \right)\right\rangle ~~~,~~~ 
R(z) = -\frac{1}{32\pi^2} \left\langle\tr 
\left( \frac{\cW^\alpha \cW_\alpha}{z-\phi} \right)\right\rangle\,,
\end{equation}
where $\phi$ is the scalar component of the chiral multiplet in the adjoint
and $\cW_\alpha$ the gaugino field.
It turns out that $R(z)$ takes values on a hyperelliptic Riemann surface
and that $T(z)$ defines a meromorphic differential on it. 
The superpotential can be evaluated in terms of period integrals
on this Riemann surface. 

Much work has been devoted to generalize the approach of \cite{Cachazo:2002ry}
to theories with different gauge groups and matter content,
see the review \cite{Argurio:2003ym}
 for a list of references.
However, most of the work so far has been
based on the presence of a chiral multiplet in the adjoint representation
of the gauge group. One exception to this is \cite{kraus},
\cite{Alday:2003gb}  where orthogonal
gauge groups with symmetric tensors and symplectic gauge groups with 
antisymmetric tensor have been considered. These theories also presented
a puzzle in how to compute correctly the effective superpotential that
took some time to resolve and understand completely \cite{Cachazo:2003kx},
\cite{Matone:2003bx}, \cite{Landsteiner:2003ph}, \cite{Intriligator:2003xs}. 
Konishi anomalies for theories with chiral spectrum and no
additional adjoints have also been studied
in \cite{Brandhuber:2003va}, \cite{Argurio:2003tp}, \cite{DiNapoli:2004rm}.

Motivated by these developments we will investigate generalized Konishi
anomaly relations for theories with unitary gauge groups and chiral matter
multiplets in the symmetric and antisymmetric representation. Contrary to
the SO/Sp case the two-index representations of unitary groups are complex,
and therefore we need two chiral multiplets in conjugate representations.
A different possibility is to combine a chiral multiplet in the antisymmetric
representation with a chiral multiplet in the conjugate symmetric 
representation and eight fundamentals to cancel the chiral anomaly. 
Not including any further matter fields there are then three models we can
consider. 

The first two only differ in the choice of symmetry of the chiral multiplets
\begin{equation}
  \label{eq:defxy}
  X^T = \epsilon X ~~,~~ Y^T = \epsilon Y\,,
\end{equation}
with $\epsilon = \pm 1$. The gauge transformations are
\begin{equation}
  \label{eq:gauge trafos}
  X \rightarrow U X U^T ~~,~~ Y \rightarrow U^* Y U^\dagger\,,
\end{equation}
where the star denotes complex conjugation and U is a unitary $N\times N$ 
matrix.

The third model has a chiral fermion spectrum, it consists of an
antisymmetric field $A$ a symmetric field $S$ and eight fundamentals $Q_f$,
where $f=1\dots 8$ denotes the flavor index.
\begin{equation}
  \label{eq:chiralspectrum}
  S^T = S ~~,~~ A^T=-A\,,
\end{equation}
with gauge transformations 
\begin{equation}
  \label{eq:chrialtrafos}
  S \rightarrow U^* S U^\dagger ~~,~~ A \rightarrow U A U^T ~~,~~ Q_f \rightarrow
U Q_f\,.
\end{equation}

For all models we will consider a polynomial tree level superpotential in 
$XY$
or $AS$ respectively. In the chiral case we will also include a coupling
between the symmetric field and the eight fundamentals.
We will discuss the classical vacua first and compute then the
generalized Konishi anomalies. We also will show that one can define
holomorphic matrix models \cite{Lazaroiu:2003vh} whose loop equations can be mapped in an large
$N$ expansion to the Konishi anomalies of the gauge theories.
Models with this spectrum and an additional chiral multiplet in the
adjoint representation have been investigated in \cite{Klemm:2003cy},
\cite{Naculich:2003cz}, \cite{Landsteiner:2003ua}.

After completion of this work \cite{Argurio:2004vu} appeared where
the non-chiral theory with antisymmetric representation is also studied.
The curve defined by the Konishi anomly there is a double cover of the
one discussed here.

\bigskip

\section{Non-chiral Models} 

In this section we will investigate the models with chiral multiplets
in mutually conjugate (anti)symmetric tensor representations and tree level 
superpotential
\begin{equation}
  \label{eq:superpot}
  W = \sum_{k=0}^{d} \frac{g_{k}}{k+1} \tr[(XY)^{k+1}]\,.
\end{equation}
\subsection{Classical Moduli Space}
As is well-known the moduli space of a supersymmetric gauge theory can
be obtained as the critical points of the superpotential modulo complexified
gauge transformations. In the case at hand it is useful to consider the
variation
\begin{equation}
  \label{eq:eom}
  X\frac{\partial W}{\partial X} = XY \sum_{k=0}^d g_k (XY)^k\,.
\end{equation}
Note that $XY$ transforms in the adjoint representation
\begin{equation}
  \label{eq:xytrafo}
  (XY) \rightarrow U (XY) U^{-1}\,.
\end{equation}
This can be used to diagonalize $XY = \mathrm{diag}(\xi_1 {\bf 1}_{N_1},\dots,
\xi_{d+1} {\bf 1}_{N_{d+1}})$ where the eigenvalues have to fulfill
\begin{equation}
  \label{eq:eigenvlaues}
  \xi W(\xi) =0\,.
\end{equation}
Note that independent of the particular choice of the couplings $g_{k}$ 
$\xi=0$ is always a possible eigenvalue in a vacuum!
We find therefore $XY = \mathrm{diag}( {\bf 0}_{N_{0}},\xi_1 {\bf 1}_{N_1},\dots, \xi_{d} {\bf 1}_{N_d})$. Since the non-vanishing eigenvalues
come from fields in the symmetric or antisymmetric representation the
gauge symmetry breaking pattern is 
\begin{equation}
  \label{eq:gaugebreaking}
  U(N) \longrightarrow \left\{
  \begin{array}{c}
   U(N_0)  \bigotimes_{i=1}^{d}\; SO(N_i)  ~~,~~ \epsilon = +1\\
   U(N_0)  \bigotimes_{i=1}^{d}\; Sp(N_i)  ~~,~~ \epsilon = -1\\
  \end{array}
  \right. \,,
\end{equation}
where\footnote{Our conventions are $SP(2)\equiv SU(2)$.} $N=\sum_{i=0}^d N_i$.

We can arrive at this conclusion also in a slightly different way.
For definiteness let us study the case of symmetric representations.
Since the gauge group is complexified we can bring any vev of the field
$Y$ to the form $Y=\mathrm{diag}({\bf 0}_{N_0}, {\bf 1}_{\tilde N})$. 
This breaks
the gauge group in a first step to $U(N_0)\otimes SO(\tilde N)$.  
Now we write
\begin{equation}
  \label{eq:Xvevi}
  X = \left(
  \begin{array}{cc}
 a&b\\ b^T& c\\
  \end{array}\right)
\end{equation}
where $a$ is a $N_0 \times N_0$ matrix, $b$ a $N_0 \times \tilde N$
matrix and $c$ a $\tilde N \times \tilde N$ matrix. 
Notice now that $a=b=0$ lies on the extended gauge orbit of $U(N_0,\C)$ which
leaves us with the matrix $c$ that transforms as a symmetric representation
of $SO(\tilde N,\C)$ and can be diagonalized to 
$c=\mathrm{diag}(\xi_1 {\bf 1}_{N_1},\dots ,\xi_d {\bf 1}_{N_d})$.
In this way we arrive at the gauge breaking pattern \erf{eq:gaugebreaking}
for $\epsilon = +1$. 
The case with antisymmetric representations works analogous.
We can now chose $Y=\mathrm{diag}({\bf 0}_{N_0},{\bf J}_{\tilde N})$
where
\begin{equation}
  \label{eq:def.J}
  {\bf J}_N =  {\bf 1}_{N / 2} \otimes i \sigma_2
~~,~~\mathrm{where}~~ i\sigma_2=\left (\begin{array}{cc} 0&1\\
-1&0\\\end{array} \right)\,.
\end{equation}
This breaks the gauge group to $U(N_0)\otimes SP(\tilde N)$.
The same arguments as before tell us that from $X$ we obtain only 
a field $c$ transforming under the antisymmetric representation
of $SP(\tilde N)$ and $c.\bf{J}_{\tilde N}$ can be brought into the
form $c = -\mathrm{diag}(\xi_1 {\bf 1}_{N_1/2},\dots ,\xi_d {\bf 1}_{N_d/2})\otimes i\sigma_2$ that breaks the group further to the second line in 
\erf{eq:gaugebreaking}. 

We thus find that the moduli space consists of isolated points given
by the eigenvalues $\xi_i$ of the adjoint valued matrix $XY$ fulfilling
$\xi_i W(\xi_i)=0$ and that the gauge breaking pattern is 
described by \erf{eq:gaugebreaking}.

\subsection{Konishi Anomalies}

Following \cite{Cachazo:2002ry} we will now discuss the chiral ring relations that follow
from generalized forms of the Konishi Anomalies. 
It is useful to remember the general chiral ring relation
\begin{equation}
  \label{eq:annil}
  \cW^{(r)}_\alpha . \Phi^{(r)} \equiv 0\,.
\end{equation}
Here $\Phi^{(r)}$ is a chiral operator transforming in the representation
$r$ of the gauge group and $\cW_\alpha$ is the chiral operator corresponding
to the gaugino field. Equation \erf{eq:annil} holds as an equivalence
relation inside the chiral ring.  
In our case this can be written more explicitly
with the help of matrix representations as
\begin{eqnarray}
  \label{eq:annilii}
  \cW_\alpha . X &\equiv & \cW_\alpha X + X \cW_\alpha^T \equiv 0\,,\\
  \cW_\alpha . Y &\equiv & -Y \cW_\alpha  - \cW_\alpha^T Y \equiv 0 \,, 
\end{eqnarray}
where $\cW_\alpha$ is in the fundamental $N\times N$ matrix representation and 
juxtaposition stands for matrix multiplication. The chiral operator
$XY$ commutes with $\cW_\alpha$ since it transforms according to the
adjoint representation. Using these relations we
see that the chiral ring of gauge and Lorentz invariant operators is spanned
by $\tr[(XY)^k],\tr[\cW^\alpha \cW_\alpha (XY)^k]$. If we
do not restrict to Lorentz singlets we also have $\tr[\cW^\alpha (XY)^k]$
these have however vanishing expectation value in a supersymmetric vacuum.
The vacuum amplitudes of the gauge and Lorentz invariant single trace operators
can be formally summed up in the generating functions
\begin{equation}
  \label{eq:defTR}
  T(z) := \left\langle\tr \left( \frac{1}{z-XY} \right)\right\rangle ~~,~~ R(z):=\left\langle \frac{-1}{32\pi^2}\tr\left(\frac{\cW^2}{z-XY}
\right)\right\rangle
\end{equation}

The generalized Konishi anomaly is the anomalous Ward identity for a
holomorphic field transformation:
\begin{equation}
{\cal O}^{(r)} \longrightarrow {\cal O}^{(r)} + \delta {\cal O}^{(r)}~~.
\end{equation}
As a relation in the chiral ring it can be written as
\begin{equation}
\label{eq:konishi}
 \left \langle \delta {\cal O}_I
\frac{\partial W}{\partial {\cal O}_I}  +
\frac{1}{32\pi^2}  {\cal W}^{\alpha}_I\,^J {\cal W}_{\alpha,J}\,^K
\frac{\partial (\delta {\cal O}_K)}{\partial {\cal O}_I } \right\rangle =0~~,
\end{equation}
where the capital indices enumerate a basis of the representation $r$.

We will investigate the generalized Konishi relations corresponding to
the field transformations:
\begin{equation}
  \label{eq:fieldtrafos}
  \delta_1 X = \frac{1}{z- XY} X ~~~,~~~ \delta_2 X = -\frac{1}{32 \pi^2} 
\frac{\cW^2}{z-XY}X\,.
\end{equation}
Let us first show that these variations are indeed symmetric or antisymmetric
respectively
\begin{eqnarray}
  \label{eq:deltaymmetry}
  (\delta_1 X)^T  &=& X^T \left(\frac{1}{z-XY}\right)^T =
 X^T\frac{1}{z}\sum_{n=0}^\infty \frac{(Y^T X^T)^n}{z^n} = \\
&=& \frac{1}{z}\sum_{n=0}^\infty \frac{(X^T Y^T)^n}{z^n} X^T =
\frac{\epsilon}{z}\sum_{n=0}^\infty \frac{(X Y)^n}{z^n} X = \epsilon\delta_1 X
\end{eqnarray}
and similarly for the variation with the $\cW^2$ insertion where one 
also has to use the chiral ring relations \erf{eq:annilii}.

We will evaluate the tree level term and the anomalous term of the Konishi
relation \erf{eq:konishi} for the variation $\delta_1 X$ now separately.
The tree level term is
\begin{eqnarray}
  \label{eq:treeleveli}
  \left\langle\tr\left( \delta_1 X \frac{\partial W(XY)}{\partial X} \right)\right\rangle = 
\left\langle\tr\left(\frac{ XY W'(XY)}{z-XY}\right)\right\rangle = 
   c(z) + T(z) z W'(z) \,. 
\end{eqnarray}
where
\begin{equation}
  \label{eq:defc}
  -\tr\left(\frac{z W'(z) - XY W'(XY)}{z-XY}\right) = c(z) \,
\end{equation}
is a polynomial of degree $d$! 

Now we evaluate the anomalous term. 
\begin{eqnarray}
  \label{eq:anomalousi}
  \left\langle
\mathrm{Tr} \left({\cal W}^{\alpha} . {\cal W}_{\alpha} . 
\frac{\partial (\delta X)}{\partial X }\right) \right\rangle &=&
\tr \left\langle\left(
\cW^{\alpha}  {\cal W}_{\alpha} \frac{\partial (\delta X)}{\partial X }
+ \cW^\alpha  \frac{\partial (\delta X)}{\partial X } \cW_\alpha^T - \right.\right.\nonumber\\ 
& &-\left.\left.
\cW_\alpha \frac{\partial (\delta X)}{\partial X }  \cW^{\alpha T} -
 \frac{\partial (\delta X)}{\partial X }\cW_\alpha^T \cW^{\alpha T}\right)\right\rangle \,,\\
\end{eqnarray} 
where $\Tr$ denotes the trace in the (anti)symmetric representation and
$\tr$ the trace in the fundamental. 
We evaluate further
\begin{eqnarray}
  \label{eq:firstterm}
\left\langle(\cW)^2_i\,^k \frac{\partial (\delta X_{km})}{\partial X_{im} }
\right\rangle &=&
\left\langle\cW^2_i\,^k \frac{z}{2}\left[\left(\frac{1}{z-XY}\right)_k\,^i
 \left(\frac{1}{z-YX}\right)^m\,_m \,+\right.\right. \nonumber\\
& &\left.\left. +\epsilon\left(\frac{1}{z-XY}\right)_k\,^m
\left(\frac{1}{z-YX}\right)^i\,_m\right]\right\rangle\nonumber\\ 
&=& -32\pi^2 \frac{z}{2} ( R(z) T(z) - \epsilon R'(z) )\,,
\end{eqnarray}
and
\begin{eqnarray}
  \label{eq:secondterm}
\left\langle(\cW^\alpha)_i\,^k \frac{\partial (\delta X_{km})}{\partial X_{il}}
\cW_{\alpha,l}\,^m 
\right\rangle &=&
\left\langle\cW^\alpha_i\,^k \frac{z}{2}\left[\left(\frac{1}{z-XY}\right)_k\,^i
 \left(\frac{1}{z-YX}\right)^l\,_m \,+\right.\right. \nonumber\\
& &\left.\left. +\epsilon\left(\frac{1}{z-XY}\right)_k\,^l
\left(\frac{1}{z-YX}\right)^i\,_m\right] \cW_{\alpha,l}\,^m
\right\rangle\nonumber\\ 
&=& 32\pi^2 \frac{z}{2} \epsilon R'(z) \,,
\end{eqnarray}
where in the last term we took the vacuum expectation value of the 
spinor values single trace operators to vanish. 
The third term in \erf{eq:anomalousi} gives the same result
as \erf{eq:secondterm} and the last term gives the same result
as \erf{eq:firstterm}

Let us now compute the Konishi anomaly relation for the variation 
$\delta_2 X$. Here it is useful to notice that
\begin{equation}
  \label{eq:deltas}
  (\delta_2 X)_{km} = (\cW^2)_k\,^l (\delta_1 X)_{lm}
\end{equation}
Using this and the previously derived relation for $\delta_1 X$ we
find for the tree level term
\begin{equation}
  \label{eq:treelevelii}
   \left\langle\tr\left( \delta_2 X \frac{\partial W_{tree}}{\partial X} \right)\right\rangle = f(z) + z W'(z) R(z)\,,
\end{equation}
where 
\begin{equation}
  \label{eq:deff}
  f(z) = \left\langle \tr\left( \frac{\cW^2
(zW'(z) - XY W'(XY))}
{32\pi^2(z-XY)} \right) \right \rangle\,.
\end{equation}
For the anomalous term we find now
\begin{equation}
  \label{eq:anomalousii}
 \left\langle \mathrm{Tr} \left({\cal W}^{\alpha} . {\cal W}_{\alpha} . 
\frac{\partial (\delta X)}{\partial X }\right) \right\rangle =
\frac{z}{2} R^2(z) 
\end{equation}

Taking these results together we obtain
\begin{eqnarray}
\label{eq:konishisi}
  \frac{z}{2}R(z)^2 - z W'(z) R(z) - f(z) &=& 0\,,\\
\label{eq:konishisii}
 z R(z) T(z) - z W'(z) T(z) -2 \epsilon z R'(z) - c(z) &=& 0\,. 
\end{eqnarray}
As is well-known in the case of adjoint representations, the equation
for $R(z)$ defines a hyperelliptic Riemann surface. We can set $R=y+W'$
and find thus
\begin{equation}
  \label{eq:hyperelliptic}
  y^2 = (W')^2 + \frac{2 f(z)}{z} \,.
\end{equation}
This equation has the somewhat unusual feature that $y$ tends to 
$\infty$ as $z$ goes to $0$. This curve is a double cover of the
$z$-plane with branchpoints at $y=0$ and a distinguished branchpoint
at $y=\infty,~ z=0$. This branchpoint is always present and 
at the fixed locus $z=0$ unless the coefficient $f_0$ in the
polynomial $f(z) = \sum_{i=0}^d f_i z^i$ vanishes.
Notice further that if in addition the coefficient $c_0$ in
$c(z) = \sum_{i=0}^d c_i z^i$ vanishes the Konishi relations
\erf{eq:konishisi}, \erf{eq:konishisii} take the same form as the ones for
$SO$ gauge group with symmetric matter ($\epsilon = +1$) or
$SP$ gauge group with antisymmetric matter ($\epsilon = -1$).
This could have been expected of course from our analysis of
the classical moduli space which showed that in a generic
 vacuum with $\xi\neq 0$ the gauge group is either $SO$ 
or $SP$. The special point $\xi=0$ in the classical moduli space
gives rise to the special fixed point at $z=0$ in the quantum
theory. 
As usual the gaugino condensates in the factor groups and their ranks are
given by period integrals on \erf{eq:hyperelliptic}
\begin{eqnarray}
  \label{eq:gauginos}
  S_i = \oint_{A_i} y dz ~~,~~ N_i = \oint_{A_i} T(z) dz \,.
\end{eqnarray}
where $A_i$ are compact cycles surrounding the cuts of $y$.

\subsection{Matrix Model}

Based on string theory considerations Dijkgraaf and Vafa \cite{DV}
conjectured that the exact effective superpotential of a confining
$\None$  supersymmetric gauge theory can be computed with the help
of a simple matrix model. This conjecture has been proved
for many different models. The perturbative part of the conjecture
can be proved using superfield techniques in perturbation theory \cite{zanon}.
A different approach is based on comparing the $1/\hn$ expansion
of the loop equations of the matrix model with the Konishi anomaly
relations of the field theory. This approach is non-perturbative in
nature although it has to be emphasized that the Konishi anomaly
relation as stated in \erf{eq:konishi} is proven to be exact only
in perturbation theory. Of course, once a one-to-one map to
the loop equations of the matrix model (and therefore
string theory) is found, one can take this as evidence for the non-perturbative
exactness of the Konishi anomaly relations.

The precise definition of the matrix model has been
worked out in \cite{Lazaroiu:2003vh} where the need of a holomorphic definition
has been emphasized.  We will follow this viewpoint here. 
The partition function of the holomorphic matrix model is given 
by
\begin{equation}
  \label{eq:mmi}
  Z = \frac{1}{|G|} \int_\Gamma d\hx d\hy e^{-\frac{1}{\kappa} 
\tr[W(\hx\hy)]}
\,,
\end{equation}
where $\hx$ and $\hy$ are the matrices corresponding to the chiral
multiplets $X,Y$ of the gauge theory. They are thus complex
$\hn\times\hn$ matrices obeying $(\hx^T, \hy^T) = \epsilon (\hx, \hy)$.
$|G|$ is a normalization factor including the volume of the gauge group
and $\Gamma$ is a suitably chosen path in the configuration space ${\cal M}$ of
the matrices $\hx, \hy$ with $\mathrm{dim}_{\R}(\Gamma) =
\mathrm{dim}_{\C}({\cal M}) $.

The matrix model action $W(\hx\hy)$ has the gauge symmetry
\begin{equation}
  \label{eq:mmgaugesym}
  \hx \rightarrow g \hx g^T ~~~,~~~  \hy \rightarrow (g^{-1})^T \hy g^{-1}\,,
\end{equation}
where $g\in GL(\hn,\C)$. Before attempting to evaluate the path integral
we therefore have to fix the gauge. Doing so we have to treat the
two cases in slightly different ways. 

Let us start with $\epsilon = +1$,
i.e. $\hx, \hy$ being symmetric. We chose the gauge $\hy = {\bf 1}_\hn$.
We could implement this via the BRST formalism. However, with this
gauge choice the ghost sector decouples from the $d\hx$ integrations.
Furthermore, it is clear that only the symmetric part of the
$gl(\hn,\C)$ valued ghost fields contribute. The gauge fixing is not
complete, gauge transformations leaving $\hy={\bf 1}_\hn$ invariant,
i.e. obeying $g g^T = 1$ survive. Therefore after this partial
gauge fixing we are left with a matrix model based on an $SO(\hn,\C)$
gauge group and a symmetric field $\hx$! At this point we can do a further
gauge fixing $\hx = \mathrm{diag}(\lambda_1,\dots,\lambda_\hn)$.
This time of course the ghost sector contributes non-trivially and
leads to the insertion of the Vandermonde determinant in 
the integral, which of course is the same one as for
the $SO$-theory with symmetric matter. We find therefore \cite{Landsteiner:2003ua}
\begin{equation}
  \label{eq:eigenvaluemodelsym}
  Z_{\epsilon=+1} = \frac{1}{|G'|} \int \prod_{k=1}^\hn d\lambda_k \prod_{m<n} |\lambda_m-\lambda_n|
e^{-\frac{\hn}{\kappa} \sum_i W(\lambda_i)}~.
\end{equation}
In the intermediate steps leading to this formula we have to chose 
an appropriate path $\Gamma$ in the matrix configuration space
including the ghosts such that all the integrals converge and $\lambda_i\in \R$.
This is by now well-studied in many examples and we therefore do not
give any more details. The interested reader is instead referred to
\cite{Lazaroiu:2003vh} and chapter 6.1 in \cite{Klemm:2003cy}. 

At this point one might conclude that the loop equation for our model
is the same as the one for the $SO$ model with symmetric matter and
that therefore a relation to the Konishi anomalies of the field
theory can not be established or could at most be established for the
vacua with $\xi\neq 0$. This would be wrong for the following reason.
The loop equations in the eigenvalue representation for the $SO$ model
follow from the identity
\begin{equation}
  \label{eq:somodel}
  0=\int \prod_{k=1}^\hn d\lambda_k \left[\sum_r
\frac{\partial}{\partial \lambda_r} \left(\frac{1}{z-\lambda_r}
\prod_{m<n} |\lambda_m-\lambda_n|
e^{-\frac{1}{\kappa} \sum_i W(\lambda_i)}\right)\right]~.
\end{equation}
It is the insertion of $\frac{1}{z-\lambda_r}$ that is problematic.
It corresponds to the variation $\delta \hx = \frac{1}{z-\hx}$.
This variation is of course valid in the model based on the $SO$
gauge group. In our model we should however not forget that the
underlying gauge symmetry is $GL(\hn,\C)$ and that $\hx$ is a 
symmetric two-tensor under this symmetry. This reasoning shows that
such an insertion is not gauge invariant in our model. Rather than
\erf{eq:somodel} the underlying gauge symmetry instructs us to
use the identity
\begin{equation}
  \label{eq:unmodel}
  0=\int \prod_{k=1}^\hn d\lambda_k \left[\sum_r
\frac{\partial}{\partial \lambda_r} \left(\frac{\lambda_r}{z-\lambda_r}
\prod_{m<n} |\lambda_m-\lambda_n|
e^{-\frac{1}{\kappa} \sum_i W(\lambda_i)}\right)\right]~,
\end{equation}
corresponding to the variation $\delta \hx = \frac{1}{z-\hx\hy}\hx$
respecting the $GL(\hn,\C)$ gauge symmetry. More explicitly
the identity \erf{eq:unmodel} is
\begin{equation}
  \label{eq:unmodleii}
  \left\langle 
    \sum_{i=r}^\hn\left( \frac{1}{z-\lambda_r} + 
\frac{\lambda_r}{(z-\lambda_r)^2}  - \frac{1}{\kappa} \frac{\lambda_r W'(\lambda_r)}{z-\lambda_r}+ \sum_{s\neq r}\frac{\lambda_r}{z-\lambda_r}\frac{1}{\lambda_r-\lambda_s} \right)
  \right\rangle = 0
\end{equation}
Here we defined
\begin{equation}
  \label{eq:vevdef}
  \langle \sum_r f(\lambda_r)\rangle = \int \prod_{k=1}^\hn d\lambda_k\,
 \sum_r f(\lambda_r) 
\prod_{m<n} |\lambda_m-\lambda_n|
e^{-\frac{1}{\kappa} \sum_i W(\lambda_i)}\,.
\end{equation}
We further define the matrix model resolvent as
\begin{equation}
  \label{eq:resolvent+}
  \omega(z) = \kappa \sum_{i=1}^\hn \frac{1}{z-\lambda_i}\,,
\end{equation}
and a polynomial 
\begin{equation}
  \label{eq:matrixpoly}
  \hf(z) = -\kappa  \sum_{i=1}^\hn 
  \frac{zW'(z) - \lambda_i W'(\lambda_i)}{z-\lambda_i}\,.
\end{equation}
Using
\begin{equation}
 \kappa \sum_i\left( \frac{1}{z-\lambda_i} + \frac{\lambda_i}{(z-\lambda_i)^2} \right) = - z \omega'(z)
\end{equation}
and\footnote{The sums in the following formula are double sums over both
indices $i$ and $j$.}
\begin{eqnarray}
  \sum_{i\neq j} \frac{\lambda_i}{z-\lambda_i}\frac{1}{\lambda_i-\lambda_j}
&=& \frac{1}{2} \sum_{i\neq j} \frac{1}{\lambda_i-\lambda_j}\left(
\frac{\lambda_i}{z-\lambda_i}-\frac{\lambda_j}{z-\lambda_j} 
\right)=\nonumber\\  
 &= &\frac{z}{2}\left( \sum_{i, j} \frac{1}{(z-\lambda_i)(z-\lambda_j)}
- \sum_i  \frac{1}{(z-\lambda_i)^2}\right)
\end{eqnarray}
we find the loop equation
\begin{equation}
  \left\langle \frac{1}{2} z \omega(z)^2 - \frac{\kappa}{2} z
\omega'(z) - z W'(z) \omega(z) - \hf(z) \right\rangle =0\,.
\end{equation}

In the case when $\epsilon=-1$ the discussion is similar. We can
chose in a first step the gauge $\hy={\bf J}_\hn$.
This leaves us with the residual gauge group $SP(\hn,\C)$ and a matter
field $\hx$ in the symmetric representation. The second step in the
gauge fixing procedure is now to set 
$\hx{\bf J}_\hn = \mathrm{diag}(\lambda_1,
\dots,\lambda_{\hn/2})\otimes {\bf 1}_2$. 
The eigenvalue representation
of the partition function coincides therefore with the one of the
model base on $SP$ gauge groups and antisymmetric matter fields 
\cite{Intriligator:2003xs}
\begin{equation}
  \label{eq:eigenvaluemodelasym}
  Z_{\epsilon=-1} = \int \prod_{k=1}^{\hn/2} d\lambda_k \prod_{m<n} (\lambda_m-\lambda_n)^4 e^{-\frac{2}{\kappa} \sum_i W(\lambda_i)}~.
\end{equation}
The same reasoning as before leads us to consider
\begin{equation}
  \label{eq:loopii} 
0=\int \prod_{k=1}^{\hn/2} d\lambda_k \left[\sum_r
\frac{\partial}{\partial \lambda_r} \left(\frac{\lambda_r}{z-\lambda_r}
\prod_{m<n} (\lambda_m-\lambda_n)^4
e^{-\frac{2}{\kappa} \sum_i W(\lambda_i)}\right)\right]~,
\end{equation}
The resolvent is now defined as
\begin{equation}\label{eq:resolvent-}
  \omega(z) = 2 \kappa \sum_{i=1}^{\hn/2} \frac{1}{z-\lambda_i}\,,
\end{equation}
and the matrix model polynomial is defined by
\begin{equation}
  \label{eq:mmpolyii}
  \hf(z) = -2\kappa  \sum_{i=1}^\hn 
  \frac{zW'(z) - \lambda_i W'(\lambda_i)}{z-\lambda_i}\,.
\end{equation}
The analogous calculations as before lead now to the loop equation.

Concluding our results for the loop equations of the matrix models
we find that
\begin{equation}
  \label{eq:loop}
  \left\langle \frac{z}{2}  \omega(z)^2 - \epsilon \frac{z}{2}\omega'(z)
- z W'(z) \omega(z) - \hf(z) \right\rangle =0\,,
\end{equation}
where  the resolvent is defined by \erf{eq:resolvent+} or 
\erf{eq:resolvent-} respectively.

To make contact with the Konishi anomaly relations in the
field theories we expand in orders of $\kappa$ which is
equivalent to an expansion in $1/\hn$
\begin{equation}
  \label{eq:expansion}
  \left\langle \omega(z) \right\rangle = \sum_{k=0}^\infty 
\kappa^k \omega_k\,.
\end{equation}
The expansion of the loop equations is 
\begin{eqnarray}
  \frac{z}{2}\omega_0(z)^2 - z W'(z) \omega(z) - \hf(z) &=& 0~~~,~{\cal O}(0)\\
  z\omega_0(z) \omega_1(z) - \frac{z}{2}\epsilon \omega_0'(z) - 
z W'(z)\omega_1(z) &=& 0~~~,~
{\cal O}(\kappa)\,,
\end{eqnarray}
We also consider the differential operator $\delta = \sum_i N_i \frac{\partial}{\partial S_i}$ and apply it to the loop equation at {\cal O}(0) to
find
\begin{equation}
  \label{eq:deltaloop}
  z \omega_0(z) \delta\omega_0(z) -z W'(z) \delta \omega_0(z) - \delta\hf(z) =0\,.
\end{equation}
Now it is easy to see that the Konishi anomalies are formally reproduced
by the first to terms in the $1/\hn$ expansion of the matrix model
loop equations if we set
\begin{eqnarray}
  \label{eq:equivalencei}
  \omega_0(z) = R(z) &,& \delta \omega_0(z) + 4 \omega_1(z) = T(z)\\
  \label{eq:equivalenceii}
  \hf(z) = f(z) &,& \delta \hf(z) = c(z)\,.
\end{eqnarray}
We also identify the filling fractions of the matrix model with
the gaugino bilinears in the gauge theory $\kappa \hn_i=S_i$
The identification of $T(z)$ in terms of matrix model quantities
implies also the relation between the free energy $F = -\kappa^2
\log(Z)$ and the field theory superpotential
\begin{equation}
  \label{eq:weff}
  W_{eff} = \delta F_0 + 4 F_1
\end{equation}
where we expanded  $F = \sum_{k=0}^\infty \kappa^k F_k$. 
This relation has also been found for the theories in
\cite{kraus}, \cite{Naculich:2003cz}, \cite{Landsteiner:2003ua}.

\subsection{Non-Perturbative Superpotential and Normalization}

In principle the formula \erf{eq:weff} proves the relation
between the effective superpotential and the matrix model
partition function
only up to an integration constant that is independent of the
tree-level couplings $g_k$. This integration constant is commonly
taken as the Veneziano-Yankielowicz part of the gaugino superpotential.
However,  as emphasized already in \cite{DV},\cite{Cachazo:2002ry} 
the matrix model
does contain indeed all the information to compute also this non-perturbative
contribution to the superpotential. We will illustrate this in the
simplest example of a tree level superpotential consisting solely
of a mass term, $W = m \tr(XY)$.

The matrix model partition function is
\begin{equation}
  \label{eq:mmZ}  Z = \frac{K}{\mathrm{vol}(U( \hn))} \int_\Gamma\, d\hx d\hy~
e^{-m \tr(\hx \hy)}\,.
\end{equation}
We have made the dependence on the volume of the gauge group explicit
and also consider an additional normalization constant $K$. We have also
set $\kappa=1$, $F_0$ and $F_1$ are therefore identified by there
scaling in $\hn$. 

The Gaussian integration leads to\footnote{The Path $\Gamma$ in the integral
can be chosen such that $X^\dagger = Y$!}
\begin{equation}
  \label{eq:fmm}
  F = \log(K) - \log(\mathrm{vol}(U(\hn))) + \frac{1}{2}\hn (\hn+\epsilon)
\log(\frac{\pi}{m}) \,.
\end{equation}
Using
\begin{equation}
  \label{eq:logvol}
  \log(\mathrm{vol}(U\hn))) = -\frac{\hn^2}{2}\log\left(\frac{\hn}{2\pi e^{3/2}}
\right) + O(\hn^0)\,,
 \end{equation}
and writing $K = \alpha^{\hn^2} \beta^\hn$ we find
\begin{equation}
  \label{eq:logZ}
  F = - \log(Z) = -\hn^2 \log(\alpha) -\hn \log(\beta)  - \frac{\hn^2}{2}
\log\left(\frac{\hn}{2 e^{3/2} m}\right) - \frac{\epsilon}{2} \hn \log\left(
\frac{\pi}{m}\right) \,.
\end{equation}
Setting $\hn=S$ we therefore have
\begin{eqnarray}
  F_0 = -\frac{S^2}{2} \log\left(\frac{\alpha^2 S}{2 e^{3/2} m} \right)~~~,~~~
F_1 = -\frac{\epsilon}{2} S \log\left(\frac{m}{\pi \beta^{\epsilon/2}}\right)
\end{eqnarray}
Setting now $\alpha = \frac{\sqrt{2}}{\Lambda}$ and 
$\beta=\left(\frac{\Lambda}{\pi}\right)^{\epsilon/2}$  where
$\Lambda$ is interpreted as the scale of the underlying gauge theory and 
using \erf{eq:weff} we compute the superpotential
\begin{equation}
  \label{eq:VYpot}
  W_{\mathrm{eff}} = S \log\left(\frac{\Lambda^{2N-2\epsilon} m^{N+2\epsilon}}
{S^N}\right)+ N S \,, 
\end{equation}
which is the expected Veneziano-Yankielowicz superpotential with the
scale matching 
$\Lambda_{\mathrm{low}}^{3N} = \Lambda^{2N-2\epsilon} m^{N+2\epsilon}$. 

It is important to note that for higher order tree-level superpotentials
and $\epsilon =-1$ we have to take into account the effects described
in \cite{Cachazo:2003kx},\cite{Landsteiner:2003ph}, \cite{Intriligator:2003xs}. For vacua in which the gauge group is
broken to $SP(N_i)$ with an antisymmetric matter field we should
keep the corresponding $S_i$ gaugino condensates different from
zero even in the case $N_i=0$ (''$SP(0)$'' - gauge group factors). 

\section{Chiral Theory}

In this section we want to study briefly the theory with chiral fermion
spectrum. It has a chiral multiplet $A$ in the antisymmetric representation,
a chiral multiplet $S$ in the conjugate symmetric representation and
eight chiral multiplets in the fundamental representations which are needed
to cancel the chiral anomaly. The superpotential we choose is
given by
\begin{equation}
  \label{eq:treechiral}
  W = \tr[V(AS)] + \sum_{f=1}^8 Q_f S Q_f  \,,
\end{equation}
where $V(z)$ is an even polynomial of order $d+1$.
Because of the symmetry properties of $A$ and $S$ odd powers in $V$ vanish 
\begin{equation}\label{eq:evenness}
\tr[(AS)^{2n+1}]=0\,.
\end{equation}

Let us first study the classical moduli space. The equations of motion
are
\begin{eqnarray}
  \label{eq:eomchirali}
  SA V'(SA) + S Q_f &=& 0 \,,\\
  \label{eq:eomchiralii}
  V'(AS) AS &=& 0\,,\\
  \label{eq:eomchiraliii}
  S Q_f &=& 0 \,.
\end{eqnarray}
We can proceed now as in the non-chiral
theory. By complexified gauge transformations we can bring $S$ into the
form $S= \mathrm{diag}({\bf 0}_{N_0},{\bf 1}_{\tilde N})$. This 
breaks the gauge group in a first step to $ U(N_0)\otimes SO(\tilde N)$.
The fundamentals have to lie in the kernel of the matrix $S$ and 
$A$ we can divide into $N_0 \times \tilde N$ blocks
\begin{equation}
  \label{eq:AQ_F}
  Q_f = \left( 
    \begin{array}{c}
     \vec{q}_f\\ 0_{\tilde N} 
    \end{array}
\right)
~~~,~~~
A = \left( 
    \begin{array}{cc}
     a & b \\ -b^T & c
    \end{array}\right)\,,
\end{equation}
where $\vec{q}_f$ are $N_0$ dimensional row vectors, 
a is an antisymmetric $N_0 \times N_0$ matrix,  b is a $N_0 \times\tilde N$ 
matrix and $c$ is an antisymmetric $\tilde N \times \tilde N$ matrix;
$a=0$, $b=0$ and $q_f=0$ all lie on an extended gauge orbit of $U(N_0,\C)$.
This leaves $c$ which transforms as an adjoint under the $SO(\tilde N)$
residual gauge symmetry. By an $SO(N_0,\C)$ gauge transformation
we can bring it to the form $c=\mathrm{diag}(\lambda_1 {\bf 1}_{N_1}\otimes
{\bf J}_2, \dots, \lambda_{N_d} {\bf 1}_{N_d}\otimes{\bf J}_2 )$ which breaks 
the
$SO(\tilde N)$ to a product of unitary gauge groups. Of course the
non-zero $\lambda_i$ have to fulfill $V'(\lambda_i)=0$.
We find therefore the classical moduli space to consist of isolated 
points parametrized by $\lambda_i$ fulfilling $\lambda_i V'(\lambda_i)=0$.
The gauge breaking pattern in a general vacuum is
\begin{equation}
  \label{eq:chiralbreaking}
  U(N) \rightarrow U(N_0) \bigotimes_{i=1}^{d} U(N_i) ~~,~~ 
\sum_{i=1}^d 2 N_i + N_0 = N \,.
\end{equation}

The generalized Konishi Anomalies we are interested in follow from
the field transformations
\begin{eqnarray}
  \label{eq:chiral.trafos}
  \delta_1 A &=& \frac 1 2 \left( \frac{1}{z-AS}A + A \frac{1}{z+SA} \right)
= \frac{z}{z^2-(AS)^2}A \\
  \delta_2 A &=& \frac 1 2 \left( \frac{\cW^2}{32\pi^2}\frac{1}{z-AS}A +   \frac{\cW^2}{32\pi^2}A \frac{1}{z+SA} \right) = \frac{\cW^2}{32\pi^2} 
\frac{z}{z^2-(AS)^2}A \,,\\
\delta Q_f &=& \rho_{fg} \frac{1}{z-AS}Q_g \,,
\end{eqnarray}
where $\rho_{fg}$ is an arbitrary matrix in flavor space. 
The resolvents we are looking
for are
\begin{equation}
  \label{eq:chiral.resolvents}
  R(z) = -\frac{1}{32\pi^2} \left\langle \tr\left(
\frac{\cW^2}{z-(AS)}\right) \right\rangle ~~,~~
T(z) = \left\langle \tr\left(
\frac{\cW^2}{z-(AS)}\right) \right\rangle ~,
\end{equation}

\erf{eq:evenness} implies that the resolvents can also be written as
\begin{equation}
  \label{eq:chiral.resolvents.ii}
  R(z) =  -\frac{1}{32\pi^2} \left\langle \tr\left(\frac{\cW^2 z}{z^2-(AS)^2}
\right)\right\rangle
~~~,~~~
 T(z) =  \left\langle \tr\left(\frac{z}{z^2-(AS)^2}\right)\right\rangle
\end{equation}

The computation of the Konishi anomalies is now straightforward and
rather similar to the non-chiral case. For this reason we will not
give any details here.
We define the order $d-1$ polynomials even in $z$ 
\begin{equation}
  \label{eq:poly.c}
  c(z) = -\tr\left(\frac{zV'(z) - AS V'(AS)}{z^2-(AS)^2}\right) \,. 
\end{equation}
and 
\begin{equation}
  \label{eq:poly.f} 
 f (z) = \tr\left( \frac{\cW^2(zV'(z)-AS V'(AS))}{32\pi^2(z^2-(AS)^2)}\right)\,.
\end{equation}

The generalized Konishi anomalies can be written then as
\begin{eqnarray}
  \label{eq:chiral.konishi.i}
  \frac{1}{2} R(z)^2 - V'(z) R(z) -  f (z) &=&0\,,\\
  \label{eq:chiral.konishi.ii}
  T(z) R(z) - \frac{2}{z} R(z) - V'(z) T(z) -  c (z) &=& 0\,,\\
  \label{eq:chiral.konishi.iii}
  \left\langle Q_f \frac{1}{z-AS} S Q_g\right\rangle &=& \frac{1}{2}R(z) \delta_{fg}\,.
\end{eqnarray}
The relations \erf{eq:chiral.konishi.i} and \erf{eq:chiral.konishi.iii}
are precisely the same as the ones for the model with $SO$ gauge 
group and a chiral multiplet in the adjoint representation 
\cite{Ashok:2002bi}. This should be contrasted to the chiral model 
with additional adjoint in \erf{Landsteiner:2003ua} whose generalized Konishi
anomaly relations for $R(z)$ and $T(z)$ have been found to be the same
as the ones for the $SO$ model with chiral multiplet in the symmetric
representation. This implies an exact non-perturbative equivalence
between these theories. An exact non-perturbative equivalence
for the non-chiral model with antisymmetric representations
in the large $N$ limit has been discussed recently in 
\cite{Armoni:2004db}. Similar equivalences have been found 
in \cite{Argurio:2004gq} for $U(N)$ theories with matter in adjoint and
fundamental representations. 

An important conclusion can be drawn if we also consider the field transformation
\begin{equation}
  \label{eq:delta.S}
  \delta S = \frac{1}{2} \left(S\frac{1}{z-AS} + \frac{1}{z+ SA} S \right)=
 S\frac{z}{z^2-(AS)^2} \,.
\end{equation}
The corresponding Konishi anomaly is
\begin{equation}
  \label{eq:konishi.S}
  c(z) + V'(z) T(z) + \left\langle Q_f \frac{1}{z^2-(AS)^2} S Q_f \right\rangle
= R(z) T(z) + \frac 2 z R(z)\,.
\end{equation}
Let us keep formally the number of flavors arbitrary. From 
\erf{eq:chiral.konishi.iii} we find
\begin{equation}
  \label{eq:chiral.konishi.iii.trace}
  \left\langle Q_f \frac{z}{z^2-(AS)^2} S Q_f \right\rangle = \frac{N_f}{2} R(z)\,.
\end{equation}
Here we are using $\langle Q_f \frac{z}{z^2-(AS)^2}S Q_f \rangle=
\langle Q_f \frac{1}{z-(AS)}S Q_f \rangle$ which holds because
of the symmetry of  $(Q_f\otimes Q_f)_{ij}$. 
The important point is that \erf{eq:konishi.S} coincides with
\erf{eq:chiral.konishi.ii} only if $N_f=8$! So consistency of the
Konishi anomaly equations allows us to obtain the same constraint on
the number of fundamental flavors as the cancellation of the chiral
anomaly. This has already been observed in the theory with adjoint
in \cite{Landsteiner:2003ua}. 

Let us now study the corresponding matrix model. It is given by

\begin{equation}
\label{eq:chiral.matrixmodel}
 Z = \frac{1}{G} \int_\Gamma d\ha d\hs d\hq e^{-\frac{1}{\kappa}\tr[V(\ha \hs)+ \hq_f \hs \hq_f]}\,.
\end{equation}
Of course this matrix model has to be understood again in the holomorphic
context \cite{Lazaroiu:2003vh}. 

To go to an eigenvalue representation we have to possibilities. First we can
chose the gauge $S={\bf 1}_\hn$. In this case the integration over the $\hq_f$
degrees of freedom are trivial and contribute only an overall factor to the
normalization. The same is true for the integration over the ghosts that 
are needed in the gauge fixing.
This leaves the residual gauge group $SO(\hn)$ and $\ha$ that transforms
as an adjoint under this gauge group. So we end up with the eigenvalue
model for an $SO(\hn)$ matrix model and an adjoint field $\ha$ \cite{Ashok:2002bi}. 
\begin{equation}
\label{eq:chiral.eigenvalu.i}
Z = \frac{1}{G'} \int_\Gamma \prod_{i=1}^\hn d\lambda_i 
\prod_{k<l} (\lambda_k^2 - \lambda_l^2)^2 e^{-\frac{2}{\kappa} \sum_n V(\lambda_n)}
\end{equation}
In the non-chiral models we had to take care now to keep track of the
original gauge symmetry which enforced a different derivation of the loop
equations. What is then the correct Ward identity leading to the loop 
equations in the chiral model? It can of course be read off from the
transformation that lead to the Konishi anomalies. In particular for the
field $\ha$  this is
\begin{equation}
\delta \ha = \frac{z}{z^2-(\ha\hs)^2}\ha \,.
\end{equation}
The corresponding Ward identity in the eigenvalue representation is
\begin{equation}
\label{eq:chiral.wi}
0 = \frac{1}{G'} \int_\Gamma \prod_{i=1}^\hn d\lambda_i \sum_{r+1}^\hn
\frac{\partial}{\partial \lambda_r} \left(
\frac{z \lambda_r}{z^2-\lambda^2_r}
\prod_{k<l} (\lambda_k^2 - \lambda_l^2)^2 e^{-\frac{2}{\kappa} \sum_n V(\lambda_n)}
\right)\,.
\end{equation}
This is however up to the overall factor of $z$ precisely the Ward identity
that leads to the loop equation for the $SO$ model with adjoint!
Indeed we also find that the definition of the matrix model resolvent
\begin{equation}
\omega(z) = \left\langle \kappa\tr\left(\frac{z}{z^2-(\ha\hs)^2} \right)\right\rangle
 = \left\langle\kappa \sum_{i=1}^\hn \frac{z}{z^2-\lambda_i^2}\right\rangle \,,
\end{equation}
matches also the the one in the $SO$ model \cite{Ashok:2002bi}.
So we infer that indeed we find the well-known loop equation
\begin{equation}\label{eq:chiral.loop}
\left\langle\frac{1}{2} \omega(z)^2 - V'(z) \omega(z) - \frac{\kappa}{2z} \omega(z) -\tilde f(z)\right\rangle =0\,.
\end{equation}

Finally let us investigate the variations
\begin{eqnarray}
  \label{eq:last}
  \delta \hs = \hs\frac{z}{z^2-(\ha\hs)^2} ~~~,~~~ \delta \hq_f = 
\frac{\rho_{fg}}{z^2-(\ha\hs)^2}\hq_g 
\end{eqnarray}
and also
keep the number of flavors explicit. We find then
\begin{eqnarray}\label{eq:sploops}
\left\langle \frac{1}{2}\omega(z)^2 + \frac{\kappa}{2z}\omega(z) -\kappa 
\hq_f \hs
\frac{1}{z^2-(\ha\hs)^2} \hq_f - V'(z)\omega(z) - \tilde f (z)
\right\rangle =0\,,\\
\left\langle \hq_f \hs\frac{z}{z^2-(\ha\hs)^2} \hq_f \right\rangle = 
\frac{\hn_f}{2} \omega(z)\,.
\end{eqnarray}
We therefore that the two ways of deriving the loop equations lead to
the same result only if $\hn_f=2$! This mismatch in the number of flavors
between the field theory and the matrix model has al
ready appeared before
in \cite{Landsteiner:2003ua}. It therefore of course not surprising at all that we
find the same mismatch here. 

The polynomial $\tilde f(z)$ in the above equations is
\begin{equation}
  \tilde f(z) = \tr\left( \frac{z V'(z) - \ha \hs V'(\ha \hs)}{z^2-(\ha\hs)^2}\right) \,.
\end{equation}
The matrix model loop equations match the Konishi anomaly relations
in precisely the same way as in the non-chiral case. We expand
the resolvent as in \erf{eq:expansion} and find also
for the chiral theory the relations \erf{eq:equivalencei}, 
\erf{eq:equivalenceii} and \erf{eq:weff}.

\section{Conclusions}
We have investigated Konishi anomaly relations and loop equation in the
corresponding matrix models for theories with unitary gauge groups but
without basic fields in the adjoint representation. Of course a composite
adjoint field is easily constructed as a bilinear of the basic fields in
two-index tensor representations. It turns out that this composite adjoint
field can be used to define generalized Konishi anomaly relations or
loop equations that are rather similar as the ones for the models with
basic adjoint field\footnote{Composite adjoints and Konishi anomalies have
also been discussed very recently in the context of chiral quiver theories 
in \cite{DiNapoli:2004rm}}. 
In contrast to the models with additional 
adjoint field the curves we found are hyperelliptic!
In the non-chiral case we found a somewhat new
and interesting feature in the hyperelliptic curve defined by the Konishi
anomalies. There was a branch point fixed at the origin and related to
the special vacuum at $\xi=0$.  

With view of all the interesting results that have been obtained in
the last years for the models with basic adjoint fields it is clear
that there is still much work to be done in these new models.
A detailed study of the phase structure should lead to a deeper 
understanding of the physics of the special vacuum with
fixed branchpoint at the origin. A different question is
how these models are realized in string theory either using 
D5-branes wrapped on orientifolds of resolutions of generalized 
conifolds in type IIB string theory or as intersecting brane
configurations a la Hanany-Witten in the type IIA/M-theory approach.
We hope to come back to these questions in a future publication.

\end{document}